\documentclass[aps,reprint,superscriptaddress,nofootinbib]{revtex4-1}
\usepackage{graphicx,xcolor,soul}%
\pdfoutput=1
\usepackage[caption=false]{subfig}
\usepackage{amsmath}%
\usepackage{amsfonts}%
\usepackage{amssymb}%
\usepackage{graphicx}
\usepackage{mathptmx}
\usepackage{bm}
\usepackage{CJK}
\usepackage[colorlinks=true,citecolor=blue,breaklinks=true,%
urlcolor=blue,linkcolor=blue]{hyperref}

\providecommand{\U}[1]{\protect\rule{.1in}{.1in}}
\def\unit#1{\mathord{\thinspace\rm #1}}
\def\func#1{\mathop{\rm #1}\nolimits}%
\def\mathvecfont#1{\textbf{\em #1}}
\newcommand{\myvec}[1]{\mathvecfont{#1}}

\usepackage[normalem]{ulem}
\usepackage{color}

\newcommand{\ket}[1]{\left| #1 \right>} % for Dirac bras
\begin{document}
\begin{CJK*}{Bg5}{bsmi}
% Title of the article
\title{Band gap and broken chirality in single-layer and bilayer graphene}
% Abbreviated title for the page headers
%\titlerunning{Band gap and broken chirality}

\author{Anastasia Varlet}
\email{varleta@phys.ethz.ch}
\affiliation{Solid State Physics Laboratory, ETH Z\"{u}rich, 8093 Z\"{u}rich, Switzerland}
\author{Ming-Hao Liu (¼B©ú»¨)}
\affiliation{Institut f\"{u}r Theoretische Physik, Universit\"{a}t Regensburg, D-93040
Regensburg, Germany}
\author{Dominik Bischoff}
\affiliation{Solid State Physics Laboratory, ETH Z\"{u}rich, 8093 Z\"{u}rich, Switzerland}
\author{Pauline Simonet}
\affiliation{Solid State Physics Laboratory, ETH Z\"{u}rich, 8093 Z\"{u}rich, Switzerland}
\author{Takashi Taniguchi}
\affiliation{Advanced Materials Laboratory, National Institute for Materials Science, 1-1
Namiki, Tsukuba 305-0044, Japan}
\author{Kenji Watanabe}
\affiliation{Advanced Materials Laboratory, National Institute for Materials Science, 1-1
Namiki, Tsukuba 305-0044, Japan}
\author{Klaus Richter}
\affiliation{Institut f\"{u}r Theoretische Physik, Universit\"{a}t Regensburg, D-93040
Regensburg, Germany}
\author{Thomas Ihn}
\affiliation{Solid State Physics Laboratory, ETH Z\"{u}rich, 8093 Z\"{u}rich, Switzerland}
\author{Klaus Ensslin}
\affiliation{Solid State Physics Laboratory, ETH Z\"{u}rich, 8093 Z\"{u}rich, Switzerland}

% Please select about four verbal keywords for your manuscript.

\date{\today}

\begin{abstract}
Chirality is one of the key features governing the electronic properties of single- and bilayer graphene: the basics of this concept and its consequences on transport are presented in this review. By breaking the inversion symmetry, a band gap can be opened in the band structures of both systems at the $K$-point. This leads to interesting consequences for the pseudospin and, therefore, for the chirality. These consequences can be accessed by investigating the evolution of the Berry phase in such systems. Experimental observations of Fabry-P\'erot interference in a dual-gated bilayer graphene device are finally presented and are used to illustrate the role played by the band gap on the evolution of the pseudospin. The presented results can be attributed to the breaking of the chirality in the energy range close to the gap.
\end{abstract}

\maketitle   % please do not remove
\end{CJK*}

\section{Introduction}

Experimentally isolated in 2004 \cite{novoselov2004}, single-layer graphene (SLG) consists of a layer of carbon atoms, arranged in a honeycomb pattern. Its unit cell is defined by two carbon atoms, usually referred to as $A$ and $B$, forming the two-atom basis of a Bravais lattice.

As predicted in $1947$ \cite{wallace1947}, it was experimentally demonstrated in $2005$ that charge carriers in graphene behave like massless Dirac Fermions \cite{novoselov2005,zhang2005}. They can be described by a two-component wavefunction $\psi$ obeying:
\begin{equation}
-i \hbar v_\mathrm{F} (\bm{\sigma} \cdot \nabla) \psi = E \psi,
\label{Dirac eq}
\end{equation}
where $v_\mathrm{F}$ is the Fermi velocity and $\bm{\sigma} = (\sigma_\mathrm{x},\sigma_\mathrm{y})$ is a vector of two Pauli matrices. Here, the analogy with quantum electrodynamics can be made by realizing that the two sublattices $A$ and $B$ play the role of spin-up and spin-down and that $\boldsymbol{\sigma}$ is not the spin but the \textit{pseudospin} operator. The direction of motion is coupled to the pseudospin orientation, as one can see from Eq.~\eqref{Dirac eq}, a property denoted as \textit{chirality}. The chirality of charge carriers has important consequences for transport. It is responsible for a Berry phase of $\pi$ \cite{novoselov2005,zhang2005} and the suppression of backward scattering \cite{Katsnelson2006,Young2009}. These two effects are the basis of the Klein paradox \cite{Katsnelson2006,Klein1929}.

Bilayer graphene (BLG) also exhibits chiral charge carriers. However, instead of following the Klein physics, BLG exhibits anti-Klein properties, due to a Berry phase of $2\pi$ \cite{Katsnelson2006,novoselov2006}. The concept of chirality in both single- and bilayer graphene and its presence in interference experiments is the focus of this review.

In Section~\ref{section_2}, we introduce the concept of chirality in SLG and BLG, and lay an emphasis on illustrating this concept, together with the concept of Berry's phase. We then focus on several interference experiments, where signatures of the chirality were successfully observed. In Section~\ref{section_3}, we consider SLG and BLG systems in which the inversion symmetry has been lifted and a band gap has been opened in the band structure. We show that this lifting results in strong out-of-plane perturbations of the pseudospin in $k$-space in the energy range close to the gap, but that the pseudospin orientation is further restored to its original state at higher energies. Finally, we focus on dual-gated BLG in Section~\ref{section_4} and present an experiment allowing for the observation of the consequences of the opening of a band gap on the chirality, probed in a Fabry-P\'erot interferometer geometry.

\section{Pseudospin and chirality}
\label{section_2}
Charge carriers in graphene are \textit{chiral}. This means that there is a handedness of their states because their pseudospin is locked to the direction of motion, giving rise to interesting tunneling properties that we review in this section.

\subsection{Pseudospin motion in SLG and BLG}
In graphene, low-energy charge carriers live around two inequivalent points in momentum space, $K$ and $K'$, called ``valleys''. In the vicinity of the $K$-point, single- and bilayer graphene can be described by the Hamiltonians \cite{slonczewski1958,DiVincenzo1984}
\begin{align}
H_\mathrm{SLG} &= v_\mathrm{F}
\begin{pmatrix}
0 & \pi^{\dag}\\
\pi & 0
\end{pmatrix},
&
H_\mathrm{BLG} &= \frac{-1}{2m^{\ast}}
\begin{pmatrix}
0 & (\pi^{\dag})^{2}\\
\pi^{2} & 0
\end{pmatrix},
\label{H}
\end{align}
respectively, where $\pi = p_\mathrm{x} + ip_\mathrm{y}$ is the momentum operator and $m^{\ast}$ is the effective mass. The above effective Hamiltonians are related to the tight-binding models through $v_\mathrm{F}=3\gamma_0a/2\hbar$ with $\gamma_0$ and $a$ the intralayer nearest neighbor hopping energy and distance, respectively, and $m^{\ast}=\gamma_1/2v_\mathrm{F}^2$ with $\gamma_1$ the interlayer nearest neighbor hopping energy.

\begin{figure}[h]
\subfloat[\label{fig lattice SLG}]{
\includegraphics[height=3.8cm]{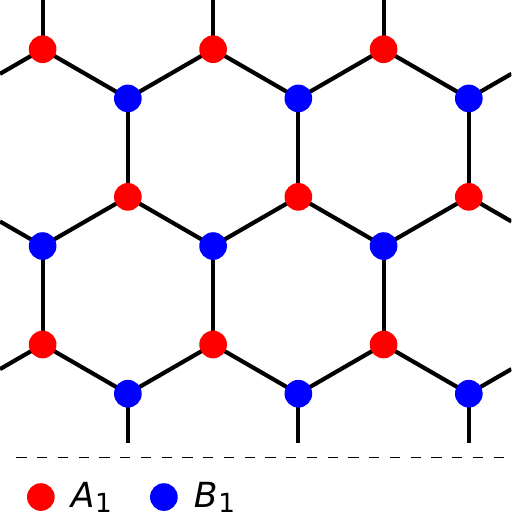}}\hfill
\subfloat[\label{fig lattice BLG}]{
\includegraphics[height=3.8cm]{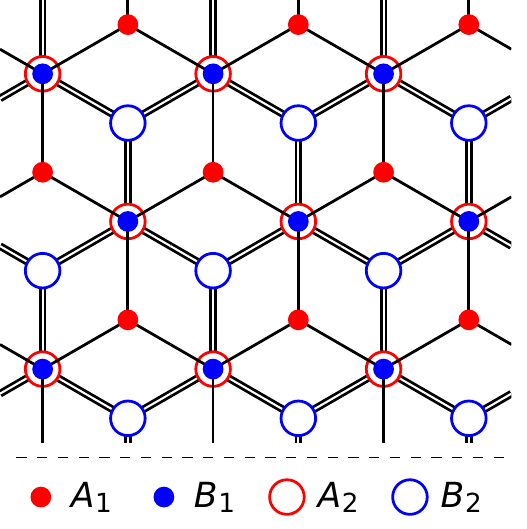}}

\subfloat[]{
\includegraphics[width=0.5\columnwidth]{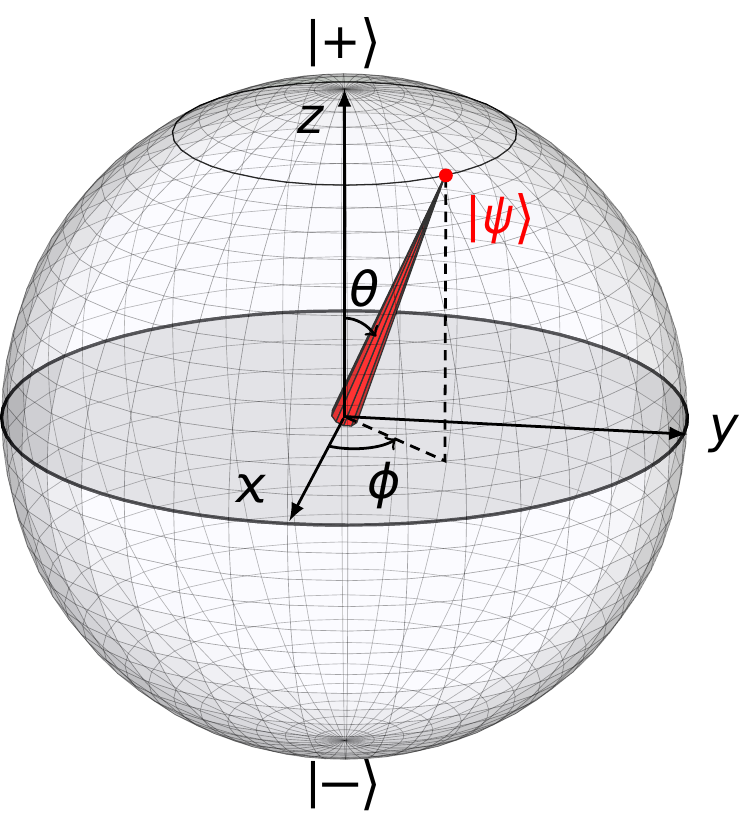}}
\subfloat[]{
\includegraphics[width=0.5\columnwidth]{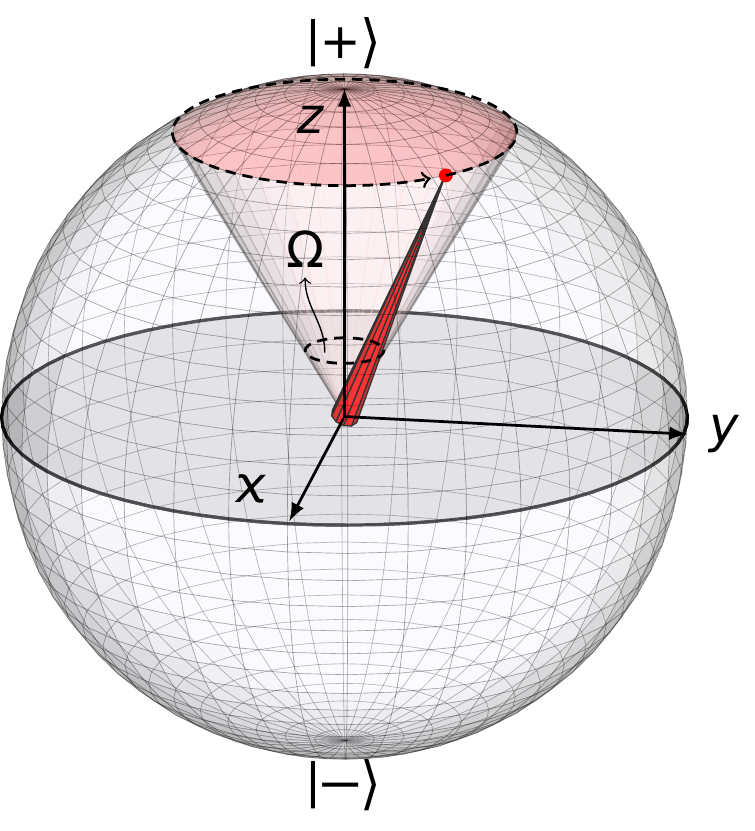}}
\caption{Lattice orientation of (a) SLG and (b) BLG in Bernal stacking. (c) A quantum state $\ket{\psi}$ and its associated polarization vector (red needle) represented on the Bloch sphere. (d) When the polarization vector undergoes an adiabatic evolution at constant polar angle $\theta$, it encloses a solid angle $\Omega$, which describes a portion of the sphere here highlighted by a red shading.}
\label{Bloch sphere}
\end{figure}

\begin{figure}
\subfloat[]{
\includegraphics[width=\columnwidth]{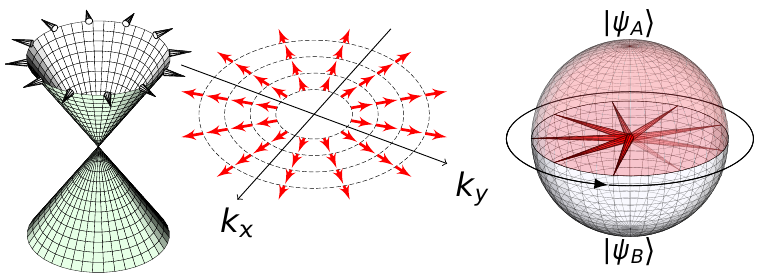}}

\subfloat[]{
\includegraphics[width=\columnwidth]{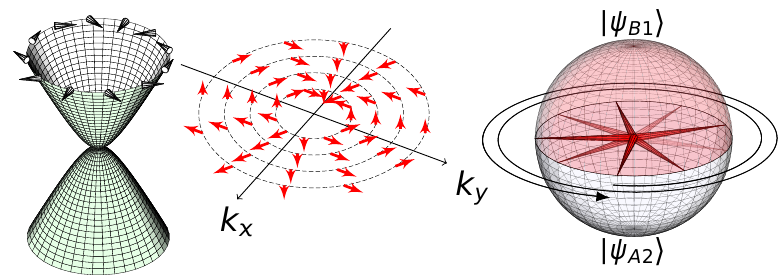}}
\caption{(a) Left: Band structure of SLG at the $K$-point with its pseudospin orientation at constant energy in the conduction band. Middle: The pseudospin orientation in $k$-space is momentum-dependent but energy-independent. Right: The pseudospin motion on the Bloch sphere when the momentum is rotated by $2\pi$. The enclosed portion of sphere is highlighted with a red shading. (b) Same content as (a) but for BLG.}
\label{pseudospin motion SLG BLG}
\end{figure}

The Hamiltonians \eqref{H} act on the spinors $(\psi_\mathrm{A},\psi_\mathrm{B})^{T}$ and $(\psi_\mathrm{B_{1}},\psi_\mathrm{A_{2}})^{T}$, respectively, where $A$ and $B$ are the two inequivalent carbon sites of a single-layer graphene and the indices $1$ and $2$ refer to the top and bottom layers of bilayer graphene; see Figs.~\ref{fig lattice SLG} and \ref{fig lattice BLG}. This gives rise to the band structure shown in Fig.~\ref{pseudospin motion SLG BLG}, left panels. Note however that the Hamiltonian \eqref{H} is a simplified two-band version, well-suited to describe the system at low energy. The full description would give rise to $4$ bands: two lower bands touching at the $K$-point (seen in the figure) and two split bands, split away from the lower ones by the energy $0.39$~eV, not represented in the figure. In the following, we consider that we are in the low energy range, where only the lower bands are filled, as shown in Fig.~\ref{pseudospin motion SLG BLG}.

The associated eigenstates are:
\begin{align}
\ket{\psi_\mathrm{SLG}^{\pm}} &= \frac{1}{\sqrt{2}} \begin{pmatrix} \pm e^{-i\phi} \\ 1 \end{pmatrix},
&
\ket{\psi_\mathrm{BLG}^{\pm}} &= \frac{1}{\sqrt{2}} \begin{pmatrix} \pm e^{-2i\phi} \\ 1 \end{pmatrix},
\label{eigenkets}
\end{align}
where the $\pm$ signs are related to the two eigenenergies $E = \pm\hbar v_\mathrm{F} k$ and $\phi =\arg {(k_\mathrm{x}+ik_\mathrm{y})}$ characterizes the direction of the wave vector $\myvec{k}=(k_\mathrm{x},k_\mathrm{y})$ measured from the $K$-point. To better visualize the details of the motion of the pseudospin which is closely related to the Berry phase, we next introduce the polarization vector $\myvec{P}$ as a convenient quantity.

In quantum mechanics, the polarization vector of a spin-1/2 quantum state $\vert\psi\rangle =(e^{-i\phi }\cos\frac{\theta}{2},\sin\frac{\theta}{2})^{T}$ is given by the expectation values of the Pauli matrices
\begin{equation}
\myvec{P}=
\begin{pmatrix}
\langle \psi \vert \sigma_\mathrm{x} \vert \psi \rangle  \\
\langle \psi \vert \sigma_\mathrm{y} \vert \psi \rangle  \\
\langle \psi \vert \sigma_\mathrm{z} \vert \psi \rangle
\end{pmatrix}
=
\begin{pmatrix}
\sin\theta\cos\phi \\
\sin\theta\sin\phi \\
\cos\theta
\end{pmatrix},
\label{P}
\end{equation}
where $\theta \in [0,\pi]$ is the polar angle and $\phi \in [0,2\pi]$ is the azimuthal angle \cite{ihn}. This vector of length $1$ describes the spin orientation of $\vert\psi\rangle$ and can conveniently be represented on the \textit{Bloch sphere} \cite{Bloch1946}. Represented on such a Bloch sphere, the quantum state $\vert \psi\rangle$, which is a superposition of the two basis states $\vert -\rangle =(0,1)^{T}$ and $\vert +\rangle = (1,0)^{T}$, together with its polarization vector are sketched in Fig.~\ref{Bloch sphere}(c).

In the case of SLG and BLG, the polarization vector can be calculated by replacing $\vert\psi\rangle$ in Eq.~\eqref{P} with the eigenstates from Eq.~\eqref{eigenkets}, leading to
\begin{align}
\myvec{P}_\mathrm{SLG} =
\begin{pmatrix}
\cos{\phi} \\
\sin{\phi} \\
0
\end{pmatrix},
&&
\myvec{P}_\mathrm{BLG} =
\begin{pmatrix}
\cos{2\phi} \\
\sin{2\phi} \\
0
\end{pmatrix},
\label{P SLG BLG}
\end{align}
which describe, instead of the real spin, the pseudospin direction of $\vert\psi_\mathrm{SLG}\rangle$ and $\vert\psi_\mathrm{BLG}\rangle$, respectively, and are particularly convenient for visualizing the motion of the pseudospin when the momentum rotates. The polarization vectors of Eq.~\eqref{P SLG BLG} have no $z$-component, due to the absence of diagonal terms in the Hamiltonians in Eq.~\eqref{H}. Thus both $\myvec{P}_\mathrm{SLG}$ and $\myvec{P}_\mathrm{BLG}$ depict a pseudospin restricted to the equatorial plane of the Bloch sphere ($\theta = \pi/2$), as sketched in Fig.~\ref{pseudospin motion SLG BLG}.

As indicated in Eq.~\eqref{P SLG BLG}, the pseudospin for the case of SLG rotates as fast as its wave vector [see Fig.~\ref{pseudospin motion SLG BLG}(a)], while for BLG it winds twice as fast [see Fig.~\ref{pseudospin motion SLG BLG}(b)]. In both cases, the process is only momentum-dependent but energy-independent. This is highlighted by the different constant energy cuts in each dispersion [middle panels of Figs.~\ref{pseudospin motion SLG BLG}(a) and \ref{pseudospin motion SLG BLG}(b)], for the case of the conduction band. For the valence band, on the other hand, the pseudospin is inverted with respect to the $K$ point. This is referred to as a chirality $+1$ for electrons (the pseudospin is always parallel to the wave vector) and $-1$ for holes (the pseudospin is always anti-parallel to the wave vector) \cite{Haldane1988}. Since the two valleys exhibit opposite handedness, the reverse situation happens around the $K'$ point, with chirality $-1$ for the electrons and $+1$ for the holes.

An important quantity closely related to the pseudospin motion is the Berry phase \cite{Berry1984}. The Berry phase, also known as a ``geometrical phase'' or ``Pancharatnam phase'', is a phase acquired by a system during an adiabatic cyclic evolution. Upon adiabatic evolution along a closed path, the polarization vector defines a portion of the Bloch sphere and subtends a solid angle $\Omega$. An exemplary motion is shown in Fig.~\ref{Bloch sphere}(d), where the polarization vector evolves along a circle of constant latitude at polar angle $\theta$. In this case, the solid angle is given by $\Omega = \int_{0}^{\theta} \int_{0}^{2\pi} \sin{\theta} \mathrm{d}\theta\mathrm{d}\phi = 2\pi( 1- \cos{\theta})$. The Berry phase $\Phi_\mathrm{Berry}$, i.e., the loop integral of the Berry connection, was found to be related to this quantity by \cite{Berry1984,Anandan1992,xiao2010}
\begin{equation}
\Phi_\mathrm{Berry} = \frac{\Omega}{2}.
\label{Berry phase = Omega/2}
\end{equation}
The Berry phase is half the solid angle subtended by the pseudospin during its motion. In case of an evolution in the equatorial plane ($\theta = \pi/2$), as in the SLG case, the Berry phase \eqref{Berry phase = Omega/2} is equal to $\pi$. In BLG, however, since the pseudospin rotates twice as fast as the momentum, the Berry phase is equal to $2\pi$ [see right panels of Fig.~\ref{pseudospin motion SLG BLG}(a)--(b)].

Hence, the Berry phase (in units of $\pi$) is equal to the number of rotations of the pseudospin when the wave vector completes one rotation around the $\myvec{k}=0$ point. This is why the Berry phase is often called the \textit{pseudospin winding number} \cite{Park2011}. This integer number represents the \textit{degree of chirality} \cite{mccann2013}. Interestingly, this description is valid as well for multi-layer graphene. A $J$-layer graphene system can be described by the Dirac-like Hamiltonian:
\begin{align}
H_{J} = g_{J}
\begin{pmatrix}
0 & (\pi^{\dag})^{J} \\
\pi^{J} & 0
\end{pmatrix}
, &&
\begin{cases}
g_{1} = v_\mathrm{F} \\
g_{2} = v_\mathrm{F}^2/\gamma_1 \\
g_{3} = v_\mathrm{F}^3/\gamma_{1}^{2} \\
\vdots
\end{cases}.
\label{eq1}
\end{align}
The corresponding Berry phase is accordingly $J\pi$ \cite{mccann2013}.

\subsection{Transmission through a potential barrier}

\begin{figure}[b]
\subfloat[]{\includegraphics[width=1.05\columnwidth]{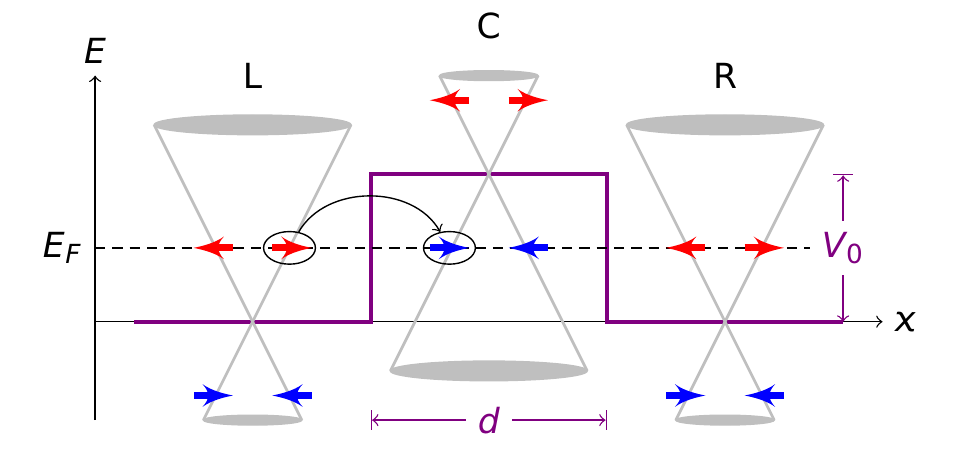}}

\subfloat[]{\includegraphics[width=1.05\columnwidth]{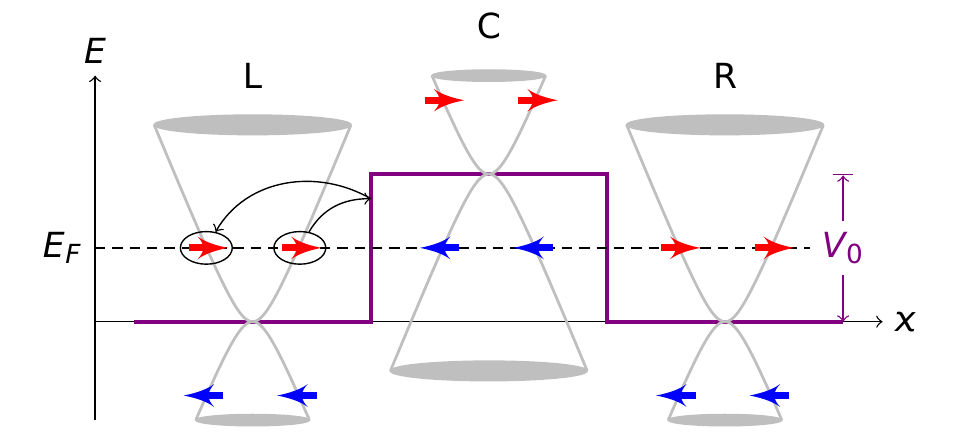}}
\caption{Schematic of a potential barrier engineered in (a) SLG and (b) BLG. The charge neutrality points are effectively shifted such that the Fermi level lies at the conduction band in the outer regions (L and R) and the valence band in the central one (C). The orientations of the pseudospins are indicated with arrows (red for electrons and blue for holes).}
\label{PotentialStepAndDispersion}
\end{figure}

The pseudospin, as described above, leads to interesting properties when considering a charge carrier incident on a potential barrier. Consider the situation depicted in Fig.~\ref{PotentialStepAndDispersion}(a): a particle, moving from left (L region) to right (R region) in SLG, is incident at energy $E_\mathrm{F}>0$ on a potential barrier (C region) of width $d$. The barrier of height $V_\mathrm{0}>E_\mathrm{F}$ is further assumed to result in a $np'n$ junction (the prime referring to the C region), shifting the charge neutrality point as schematically displayed in Fig.~\ref{PotentialStepAndDispersion}(a). If the electron propagates ballistically in the regions away from the edges of the barrier potential, such a geometry resembles a Fabry-P\'erot interferometer.

Focusing on the left interface of the barrier, we can see that a right moving electron (belonging to the right branch of the conduction band in the region L) will find a perfect match on the other side of the interface (C region), since a right moving hole exhibits the same pseudospin orientation, as highlighted in Fig.~\ref{PotentialStepAndDispersion}(a) with the encircled pseudospins. Such a situation implies a high transmission probability through the barrier. The exact solution to this problem has been investigated by Katsnelson et al.~\cite{Katsnelson2006} in 2006 and described in detail in Ref.~\cite{Tudorovskiy2012}. For an electron wave of energy $E$ incident on a barrier of height $V_\mathrm{0}>0$, with translational invariance along the transverse direction $y$, and holes populating the C region ($V_\mathrm{0} > E + \hbar v_\mathrm{F} |k_\mathrm{y}|$), the resulting reflection coefficient was found to be:
\begin{align}
\begin{split}
r &= 2 e^{i\varphi}\sin{(q_{x}d)}\times
\\
&\frac{\sin{\varphi} - ss'\sin{\vartheta}}{ss'[e^{-iq_{x}d}\cos{(\varphi + \vartheta)} + e^{iq_{x}d}\cos{(\varphi - \vartheta)}] - 2i\sin{q_{x}d}},
\end{split}
\label{reflect}
\end{align}
where $s=\func{sgn}(E)$ and $s'=\func{sgn}(E-V_\mathrm{0})$ are the sign functions, $\varphi$ and $\vartheta$ are the incident and refractive angles, and $q$ is the transmitted wave vector \cite{Katsnelson2006,Tudorovskiy2012}. This enables the calculation of the transmission probability, given by $T = 1 - |r|^2$. From this expression, one finds that normal incidence always leads to full transparency of the barrier ($r\vert_{\varphi=\vartheta=0}=0$), known as the Klein tunneling in SLG.

\begin{figure}
\subfloat[]{\includegraphics[width=0.5\columnwidth]{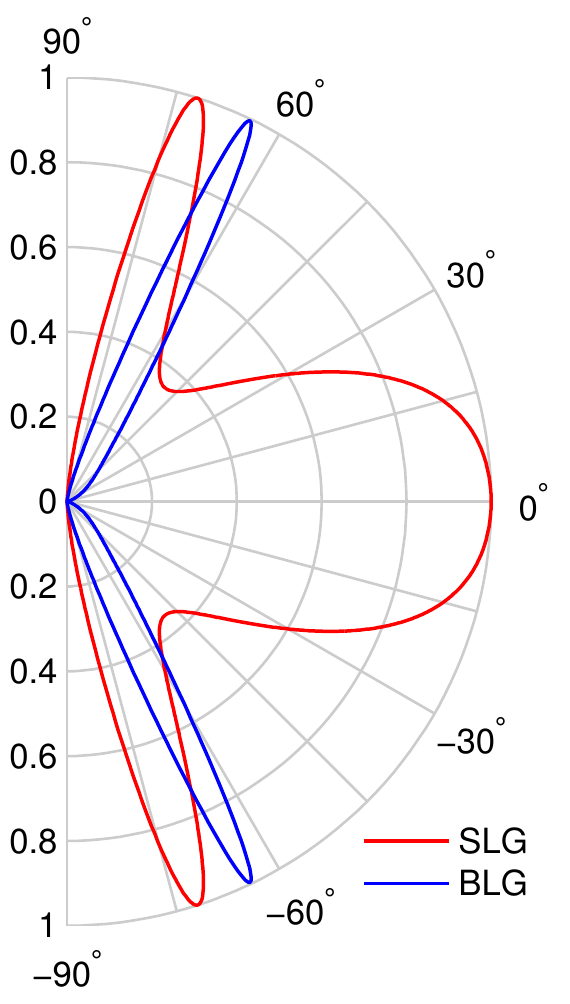}}
\subfloat[]{\includegraphics[width=0.5\columnwidth]{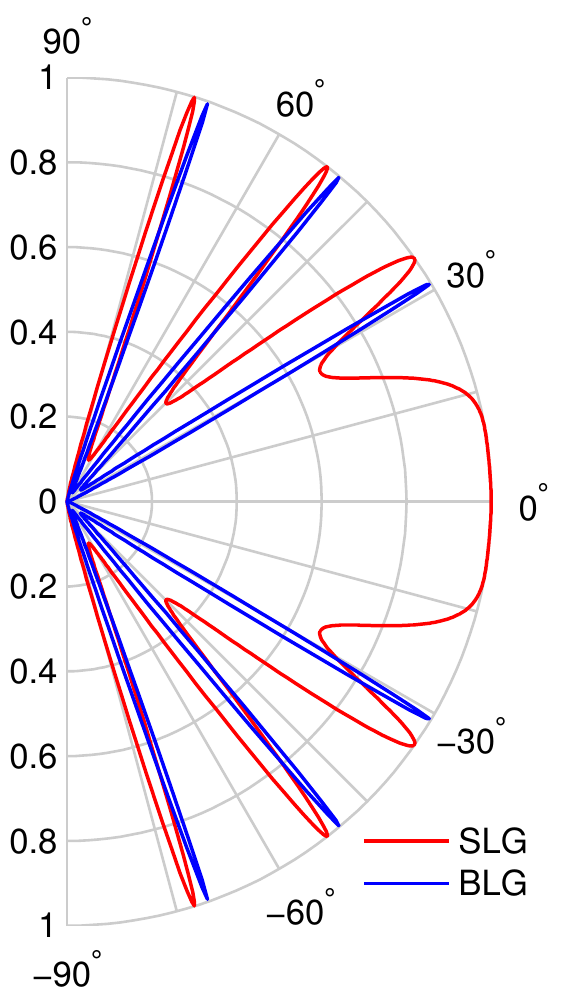}}
\caption{Calculated transmission probability $T$ as a function of the incident angle $\phi$ across a \textit{np'n} junction with fixed carrier density $n_\mathrm{out}=5\times 10^{11}\unit{cm}^{-2}$ outside the barrier and $n_\mathrm{in}=-3\times 10^{12}\unit{cm}^{-2}$ inside the barrier for SLG (red lines) and gapless BLG (blue lines) with barrier thicknesses of (a) $d=100\unit{nm}$ and (b) $d=400\unit{nm}$. The given densities $n_\mathrm{out}$ and $n_\mathrm{in}$ correspond to the incident energy $E_\mathrm{F}\approx 80\unit{meV}$ and barrier height $V_\mathrm{0}\approx 280\unit{meV}$ for SLG, and $E_F\approx 16\unit{meV},V_0\approx 100\unit{meV}$ for gapless BLG.}
\label{angle-dep}
\end{figure}

Experimentally, such a graphene \textit{np'n} junction can be realized by electrically controlling the carrier density inside and outside the barrier region. Considering densities $n_\mathrm{out}=5\times 10^{11}\unit{cm}^{-2}$ outside the barrier region and $n_\mathrm{in}=-3\times 10^{12}\unit{cm}^{-2}$ within, we show two examples of $T(\varphi)$ with barrier thicknesses of $d=100\unit{nm},400\unit{nm}$ in Fig.~\ref{angle-dep} (red curves), based on Eq.~\eqref{reflect}. These density values correspond to an incident energy of $E_\mathrm{F} \approx 80\unit{meV}$ and a barrier height of $V_\mathrm{0} \approx 280\unit{meV}$. In these plots, we can see that the transmission probability at normal incidence is one, remains very high for small incident angles, and then exhibits additional full-transmission resonances at finite angles (sometimes) called `magic angles'.

The same problem in the case of BLG was also treated in \cite{Katsnelson2006,Tudorovskiy2012} in the absence of the gap. The angle dependence of $T$ can be similarly obtained by solving the transmission problem. For a direct comparison to SLG, we consider the same $n_\mathrm{in}$ and $n_\mathrm{out}$ that lead to $E_F\approx 16\unit{meV},V_0\approx 100\unit{meV}$ and show again two examples with barrier thicknesses $d=100\unit{nm},400\unit{nm}$ in Fig.~\ref{angle-dep} (blue curves). Unlike for SLG, where massless Dirac fermions are always perfectly transmitted at normal incidence, a perfect reflection\footnote{This is true, however, only when $d\gg\lambda_F$ where $\lambda_F$ is the Fermi wavelength. In the case of $d\lesssim\lambda_F$, evanescent modes can lead to finite transmission even at normal incidence.} at normal incidence is observed. This phenomenon is known as anti-Klein tunneling, and can be understood in terms of lack of pseudospin matching as sketched in Fig.~\ref{PotentialStepAndDispersion}(b).

For non-zero incidence angles, some `magic angles' appear, where the transmission increases sharply to one. The resulting conductance, which is proportional to the transmission integrated over all the incident angles, is therefore much smaller than in the SLG case. This is a property of practical interest in the sense that electrostatic barriers in BLG are then highly efficient to confine carriers. For wider barriers, the number of resonances increases quickly, as seen already by comparing the two cases shown in Fig.~\ref{angle-dep}.

One should mention at this stage that these results of $T(\varphi)$, based on the Dirac equation, can be reproduced by the tight-binding-model-based Green's function approach, which can easily handle arbitrarily shaped barriers and also allows for more general band structures \cite{Liu2012}. This will be used in later sections to implement the band gap in BLG.

\subsection{Signatures of chirality in interference experiments\label{sec signatures of chirality}}

As explained above, the transmission across a barrier in graphene exhibits a unique behavior due to the linear dispersion and the chirality as above-defined, in contrast to non-chiral particles \cite{Katsnelson2006}. However, a direct measurement of the angle-dependent transmission $T(\varphi)$ was so far not accessible \cite{Sutar2012,Rahman2015}, since transport experiments usually measure the total conductance $G$ involving all contributing angles. This is why a true hallmark of Klein scattering has been sought. In 2008, Shytov et al.~\cite{Shytov2008} realized that a signature of Klein tunneling in ballistic $pn'p$ Fabry-P\'erot SLG cavities should appear in the magnetic field dependence of the interference pattern.

\begin{figure}[t]
\subfloat[]{
\includegraphics[scale=0.95]{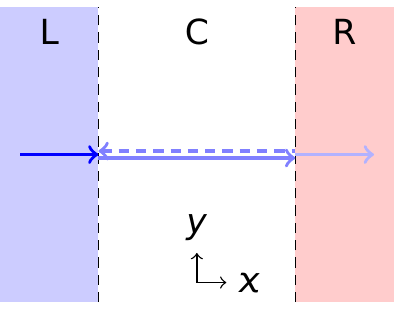}}
\hfill
\subfloat[]{
\includegraphics[scale=0.95]{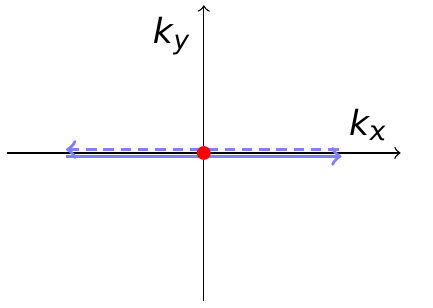}}
\caption{(a) Trajectory of a wave in real space at zero magnetic field incident from region L toward region R across a central cavity C, assuming an imperfect transmission. In C, the reflected (dashed arrow) and transmitted (solid arrow) waves are slightly offset for clarity. (b) The corresponding trajectory in momentum space with $k_y=0$ (slightly offset for clarity) encloses the origin and is equivalent to a rotation of the momentum by $2\pi$, resulting in the contribution of the Berry phase.}
\label{real and k trajectories without B}
\end{figure}

\begin{figure}[t]
\subfloat[]{
\includegraphics[scale=0.95]{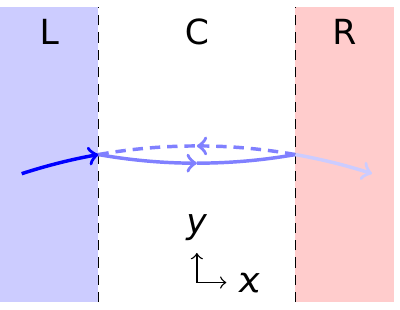}}
\hfill
\subfloat[]{
\includegraphics[scale=0.95]{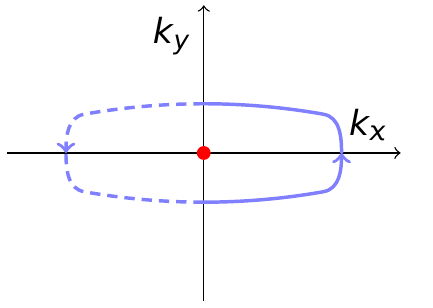}}
\caption{For SLG, the situation depicted in Fig.~\ref{real and k trajectories without B} does not exist due to the perfect transmission of Klein tunneling. Instead, a finite magnetic field is needed to obtain enough backscattering in the region C. An example of the resulting trajectory is sketched in (a). (b) The corresponding trajectory in momentum space forms a closed loop that encloses the origin. At this magnetic field value, the Berry phase of $\pi$ is picked up, leading to a phase shift in the interference pattern.}
\label{real and k trajectories with B}
\end{figure}

In the absence of magnetic field, the directly transmitted and twice reflected waves\footnote{Waves of multiple reflections also contribute to the Fabry-P\'erot interference but are only of minor importance, even in ideally ballistic transport; see, for example, \cite{varlet_fabry_2014}.} [such as the case sketched in Fig.~\ref{real and k trajectories without B}(a)] interfere with each other with a kinetic phase difference $\Phi_{\rm WKB}$ (WKB standing for Wentzel-Kramers-Brillouin), which is proportional to the cavity size $d$ and the longitudinal component of the wave vector $k_x$. Under the influence of an external magnetic field $B$, the resulting Lorentz force bends the semiclassical electron trajectories, some of which form closed loops [such as the case sketched in Fig.~\ref{real and k trajectories with B}(a)], wrapping finite areas and therefore an Aharanov-Bohm phase $\Phi_{\rm AB}$, which can be shown (when $B$ is weak) to be $\propto B^2/k_F$, where $k_F$ is the Fermi wave vector. At the same time, $B$ also reduces $k_x$ as a consequence of the bending and the conservation of $k_F=(k_x^2+k_y^2)^{1/2}$, where $k_y$ is proportional to $B$ because of the mechanical momentum $\myvec{p}\rightarrow \myvec{p}+e\myvec{A}$ with the chosen gauge\footnote{Note that when applying the periodic boundary condition along the transverse dimension ($y$), this gauge is practically the only choice in order to keep the system $y$-independent \cite{Liu2012a}.} $\myvec{A}=(0,xB,0)$ for $\myvec{B}=\nabla\times\myvec{A}=(0,0,B)$. Thus the increase of $B$ leads to an increase of $\Phi_{\rm AB}$ but decrease of $\Phi_{\rm WKB}$, while the increase of $k_\mathrm{F}$ leads to a decrease of $\Phi_{\rm AB}$ and increase of $\Phi_{\rm WKB}$. When measuring the conductance as a function of magnetic field and density ($\propto k_F^2$), the two competing phase terms result in the Fabry-P\'erot interference fringes (each stripe corresponding to a constant $\Phi_{\rm WKB}+\Phi_{\rm AB}$) dispersing with increasing $B$ towards higher density in a parabolic pattern \cite{Young2009,Shytov2008,RamezaniMasir2010,seung2011,%
Liu2012a,rickhaus_ballistic_2013,grushina_ballistic_2013,%
Masubuchi2013,Rickhaus2015,calado2015}.

The cyclotron bending contributes to the phase in a different way. At $B=0$, the trajectory propagating at normal incidence will be fully transmitted. Thus there exists no such trajectory as the one depicted in Fig.~\ref{real and k trajectories without B}(a) which, in momentum space, would directly enclose the origin [see Fig.~\ref{real and k trajectories without B}(b)]. In the presence of a finite (and typically weak) $B$, however, closed loops such as that shown in Fig.~\ref{real and k trajectories with B}(a) can form because of finite backscattering at oblique incidence, and the corresponding momentum-space trajectories will enclose the origin as sketched in Fig.~\ref{real and k trajectories with B}(b). Due to the existence of a singularity at the $K$-point (vertex \cite{Anandan1992,Shytov2009}), the $\pi$-Berry phase is picked up. In the measurements, this will therefore lead to a shift of the oscillation pattern by half a period. Hence this $\pi$-shift constitutes a direct evidence for Klein tunneling, through the analysis of backscattering.

The observation of the phase jump has been made for SLG cavities of various qualities and widths \cite{Young2009,seung2011,rickhaus_ballistic_2013,%
grushina_ballistic_2013,Masubuchi2013,Rickhaus2015}. In 2009, Young and Kim presented their pioneering experiment and reported on the observation of conductance oscillations in a graphene bipolar heterojunction \cite{Young2009}. The cavity was electrostatically induced by the use of a narrow top gate ($\sim 20~\unit{nm}$), allowing for the observation of ballistic interference, as the mean free path was estimated to be larger than $100~\unit{nm}$. The signature of Klein tunneling and of its perfect transmission was further demonstrated by investigating the magnetic field dependence of the conductance oscillations, where the $\pi$-shift of the oscillations was seen. The specificity of the measured shift of the oscillations agreed with the theoretical prediction \cite{Shytov2008} and was later reproduced qualitatively \cite{RamezaniMasir2010} and quantitatively \cite{Liu2012a} by transport calculations.

In 2013, a big step was made in the quality of the produced devices. Rickhaus et al.~\cite{rickhaus_ballistic_2013} and Grushina et al.~\cite{grushina_ballistic_2013} reported on the use of local bottom gates in suspended devices to engineer smooth, electrostatically defined $pn$ junctions with stunning qualities, enabling for the observation of Fabry-P\'erot interference on length scales larger than $1~\unit{\mu m}$ \cite{maurand2014}. Once again, the characteristic Berry phase shift was present in the magnetic field data. In Ref.~\cite{rickhaus_ballistic_2013}, the finesse of the cavity and the resulting visibility of the oscillation pattern were studied in great detail and analyzed in view of the smoothness of the electrostatic landscape.

In the case of BLG, however, the Berry phase of $2\pi$ could not be identified in a similar way in an interferometer geometry, since a phase of $2\pi$ is equivalent to a phase of $0$. Until today, the only measurable contribution of this phase was observed in quantum Hall measurements, as experimentally demonstrated for the first time in Ref.~\cite{novoselov2006}. Moreover, an electrostatically controlled BLG \textit{pn'p} junction normally requires a dual gated cavity where the inversion symmetry is broken such that the BLG is no longer gapless. The consequence of the band gap on transport and its relation to the pseudospin are to be discussed in the following section.

\section{Band gap and peudospin}
\label{section_3}
The pseudospin in SLG and BLG is restricted to the $x$-$y$ plane as a consequence of the lattice inversion symmetry. If this symmetry is broken, the situation is different.

\subsection{Inversion symmetry breaking}
By applying a different potential to the two sublattices, the inversion symmetry can be lifted. In SLG, this can be done by aligning the flake on a hexagonal boron nitride substrate \cite{decker2011,yankowitz2012,Gorbachev2014,Lundeberg2014}. The closely similar lattice structure of graphene and hexagonal boron nitride (h-BN) results in locally aligning the atomic site A of graphene with a boron atom and its B site with a nitrogen atom (or vice-versa). The two carbon sites therefore experience different potentials, leading to the broken inversion symmetry. In BLG, setting the two layers to different potentials, using for example external gates \cite{mccann_landau-level_2006,mccann_asymmetry_2006,mccann2007,mucha-kruczynski_influence_2009}, also breaks the inversion symmetry. The experimental realization of the BLG symmetry breaking will be treated in more detail in Section~\ref{section_4}.

\begin{figure}[b]
\centering
\subfloat[]{
\includegraphics[width=0.6\columnwidth]{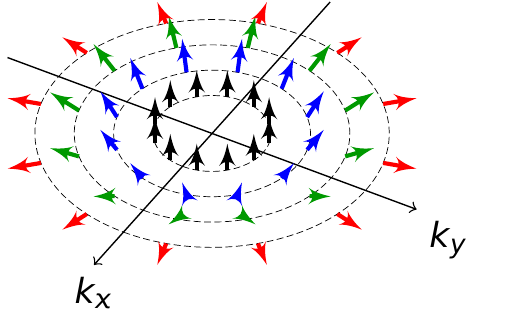}}
\subfloat[]{
\includegraphics[width=0.3\columnwidth]{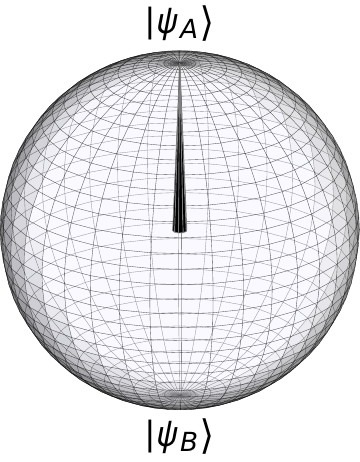}
}

\subfloat[]{
\includegraphics[width=0.3\columnwidth]{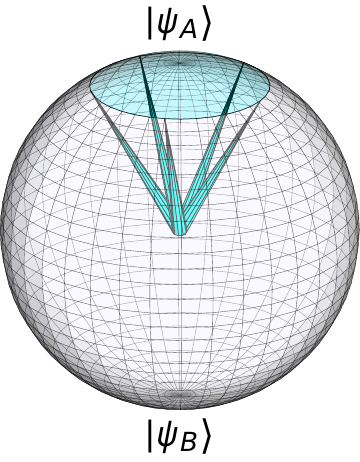}
}
\subfloat[]{
\includegraphics[width=0.3\columnwidth]{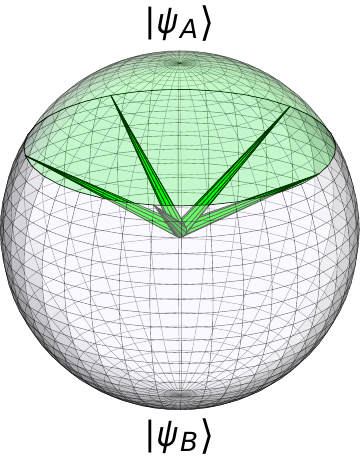}
}
\subfloat[]{
\includegraphics[width=0.3\columnwidth]{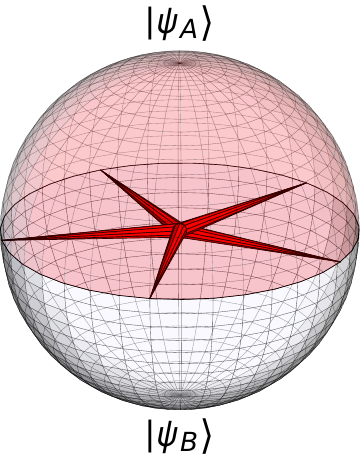}
}
\caption{(a) Pseudospin orientation along constant energy contours of the conduction band of gapped SLG (dashed circles), around the $K$-point. The process is now energy-dependent. The pseudospin motion on the Bloch sphere is shown for four cases. (b) The black needle corresponds to the black pseudospin polarization vectors close to the band gap in (a): no area is enclosed and the Berry phase is therefore $0$. (c)--(e) When going away from the gap, the $z$-component is decreased and the enclosed area grows [colored needles correspond to the pseudospin vectors sketched in (a)], until chirality is restored at higher energies and the pseudospin returns to the equatorial plane [red arrows in (a) and red needles in (e)], recovering to the Berry phase of $\pi$.}
\label{fig pseudospin with z}
\end{figure}

With the inversion asymmetry, the resulting Hamiltonians are generalized from Eq.~\eqref{H} to the form
\begin{align}
H_\mathrm{SLG}=
\begin{pmatrix}
\dfrac{u}{2} & v_{F}\pi^{\dag}\\
v_{F}\pi & -\dfrac{u}{2}
\end{pmatrix},
&&
H_\mathrm{BLG}=
\begin{pmatrix}
\dfrac{u}{2} & -\dfrac{(\pi^{\dag})^{2}}{2m^{\ast}}\\
-\dfrac{\pi^{2}}{2m^{\ast}} & -\dfrac{u}{2}
\end{pmatrix},
\label{H with u}
\end{align}
where $u$ is the asymmetry parameter \cite{mccann_landau-level_2006}.

One observes in Eq.~\eqref{H with u} that adding an asymmetry adds diagonal terms to the Hamiltonians, i.e.~a $\sigma_{z}$-component, which gives rise to a $z$-component of the polarization vector itself. In this situation, the pseudospin is not bound to the equatorial plane any more. This is shown schematically in Fig.~\ref{fig pseudospin with z}(a) for the case of the conduction band of SLG at the valley $K$. We see that for small momenta (i.e. small energies, close to the band gap), the pseudospin points completely out of plane (black arrows). On the Bloch sphere, this means that the pseudospin stays aligned with the $z$-axis and therefore encloses no area, giving rise to a zero Berry phase. This is shown in Fig.~\ref{fig pseudospin with z}(b) with the black needle. The chirality is \textit{broken}. Moving towards higher energies, the pseudospin tends to asymptotically recover its in-plane motion [red arrows in Fig.~\ref{fig pseudospin with z}(a) and red needles in Fig.~\ref{fig pseudospin with z}(e)] and therefore a $\pi$-Berry phase. The chirality is slowly restored \cite{Lundeberg2014}. In-between the two extreme cases, the pseudospin is partially $z$-polarized, leading to a Berry phase varying between $0$ and $\pi$, as shown with the blue and green arrows in Fig.~\ref{fig pseudospin with z}(a) and the blue and green needles in Fig.~\ref{fig pseudospin with z}(c)--(d).

\begin{figure}
\includegraphics[scale=1.05]{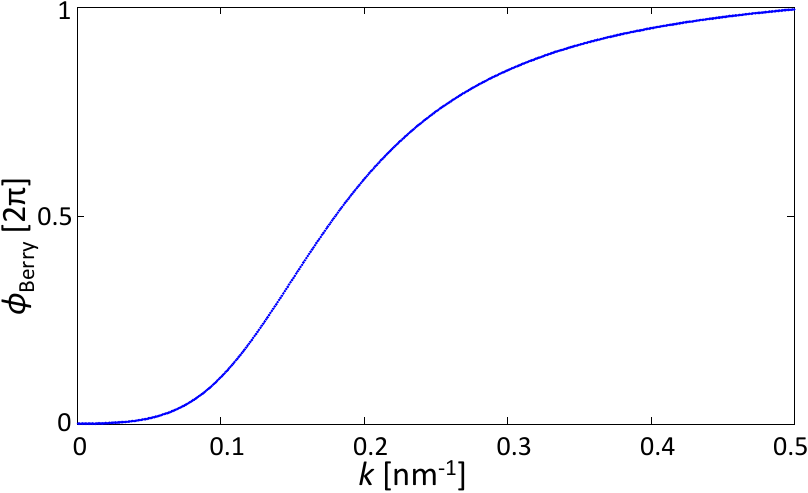}
\caption{Berry phase $\Phi_\mathrm{Berry}$ dependence (neglecting trigonal warping) on momentum $k$ for an asymmetry $u = 50\unit{meV}$ applied between top and bottom layer of BLG. As shown here, close to the gap ($k = 0$), the Berry phase is zero. Going towards higher momenta, and therefore towards higher energies, the Berry phase recovers its pristine value of $2\pi$.}
\label{Figure_PhiBerry_withGap}
\end{figure}

Exactly the same happens with BLG when a band gap is opened. Close to the edges of the valence or conduction band, the pseudospin is fully $z$-polarized, leading to a Berry phase of $0$ \cite{Li2012}. When the energy increases, the chirality is slowly restored, until the pseudospin returns to the $xy$-plane and the Berry phase is set back to $2\pi$. This is illustrated in Fig.~\ref{Figure_PhiBerry_withGap}, where the Berry phase is calculated as a function of momentum (which can be translated into an energy), and will be further explained and investigated in Section~\ref{section_4}.

\subsection{Gapped BLG \textit{pn} junction}

We next focus on the effect of the broken chirality induced by opening the band gap on the transmission probability of an electron wave incident on a BLG \textit{pn} junction. As illustrated earlier, the perfect reflection across a bipolar potential step in BLG is expected \cite{Katsnelson2006} for pristine BLG in the absence of the band gap. However, only limited literature exists addressing the question of the effect of a gap on the tunneling properties (see, for example, Refs.~\cite{Park2011,Rudner_2011}). Using the same method as Ref.~\cite{Liu2012}, we apply the Green's function method based on a tight-binding model associated with the periodic Bloch phase to illustrate the influence of the asymmetry $u$ on the angle-resolved transmission $T(\varphi)$. This asymmetry results in opening a band gap in the band structure of BLG \cite{mccann_landau-level_2006,mccann_asymmetry_2006,mccann2007,mucha-kruczynski_influence_2009}.

\begin{figure}[t]
\subfloat[]{\includegraphics[scale=0.8]{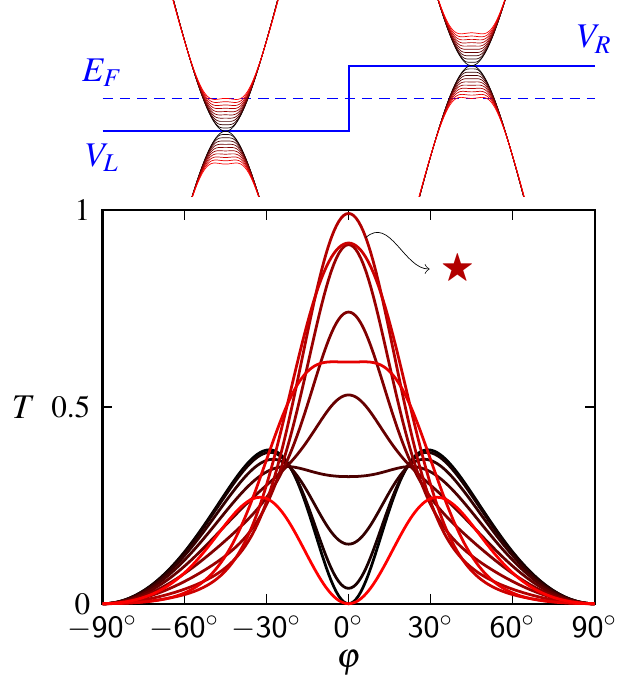}}
\subfloat[]{\includegraphics[scale=0.8]{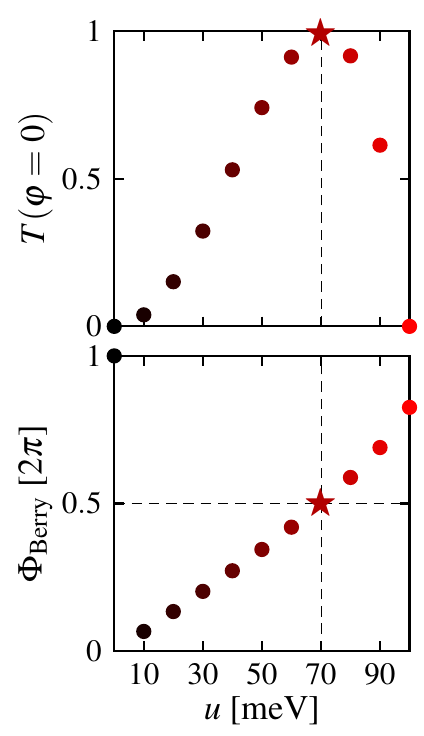}}
\caption{(a) Top: An ideal BLG \textit{np} junction with step height $V_R - V_L = 100\unit{meV}$ and $E_F-V_L = 50\unit{meV}$ with the asymmetry varying from $u=0$ (black curves) to $u=100\unit{meV}$ (red curves). Bottom: Angle-resolved transmission $T(\varphi)$ with the colors corresponding to those of the band structures shown in the top panel. (b) Top: Transmission at normal incidence $T(\varphi=0)$ from (a) as a function of $u$. Bottom: The corresponding Berry phase. The data points at $u=70\unit{meV}$ represented by $\bigstar$ correspond to $T(\phi=0) \to 1$ (top) and $\Phi_\mathrm{Berry} \to \pi$ (bottom), i.e., a revival of the Klein tunneling in BLG.}
\label{GapOpening}
\end{figure}

We consider transport through an ideally sharp \textit{np} junction\footnote{Note that an atomically sharp \textit{np} junction is expected to induce intervalley scattering, which is not our main focus here. Numerically, such scattering leads to imperfection of $T$ at $\phi=0$ for the case of SLG, and for the case of BLG with $\Phi_\mathrm{Berry}\rightarrow\pi$, as can be noted in Fig.~\ref{GapOpening}(a); see the curve marked by $\bigstar$.} with a step height of $100\unit{meV}$ at fixed Fermi energy $E_F = 50\unit{meV}$ as shown in the top panel of Fig.~\ref{GapOpening}(a), where the asymmetry parameter $u$ is varied from $0$ (black curves) to $100\unit{meV}$ (red curves). The resulting $T(\varphi)$ curves for various values of $u$ in this range are shown in the bottom panel of Fig.~\ref{GapOpening}(a), where one observes that the transmission function changes drastically. Starting from $u = 0$ [black curve with $T(\varphi=0) = 0$], we observe that increasing $u$ breaks the anti-Klein tunneling, and $T(\varphi=0)$ increases slowly, until reaching perfect transmission at normal incidence (curve marked by $\bigstar$). The $T(\varphi=0)$ values and the corresponding Berry phase as a function of $u$ are respectively shown in the top and bottom panel of Fig.~\ref{GapOpening}(b), where $T(\varphi=0)=1$ and $\Phi_\mathrm{Berry}=\pi$ matching each other at $u=70\unit{meV}$ can be seen. This implies that, under certain circumstances, a revival of the \emph{Klein tunneling in BLG} is possible, by manipulating the gap-controlled Berry phase.

By further increasing $u$, the Berry phase $\Phi_\mathrm{Berry}$ continues to increase toward $2\pi$ while the normal incident transmission $T(\varphi=0)$ decreases rapidly toward zero, until the anti-Klein tunneling behavior is recovered. In view of the transition shown and discussed above, the strength of BLG lies in the following fact: the tunability of bilayer graphene allows not only to conveniently open (and control) a band gap in its dispersion, but also to tune its chirality, enabling a possible switching between BLG-like and SLG-like transport characteristics. In the following, we present an experiment illustrating how the chirality in a gapped BLG system can be broken.

\section{Interference experiment with gapped BLG}
\label{section_4}

As mentioned earlier in Sec.~\ref{sec signatures of chirality}, there is no report in the literature showing explicit signatures of the $2\pi$-Berry phase in BLG interferometers, as there is in SLG. However, as described in the previous section, once a band gap is opened, the chirality of BLG is changed and a finite transmission at normal incidence, associated to a finite Berry phase, occurs. In the following, we investigate the chiral properties of BLG via measurements performed on a dual-gated BLG Fabry-P\'erot (FP) interferometer. Parts of the data presented below were published in Ref.~\cite{varlet_fabry_2014}.

\begin{figure}[b]
\begin{center}
\subfloat[]{
\includegraphics[width=\columnwidth]{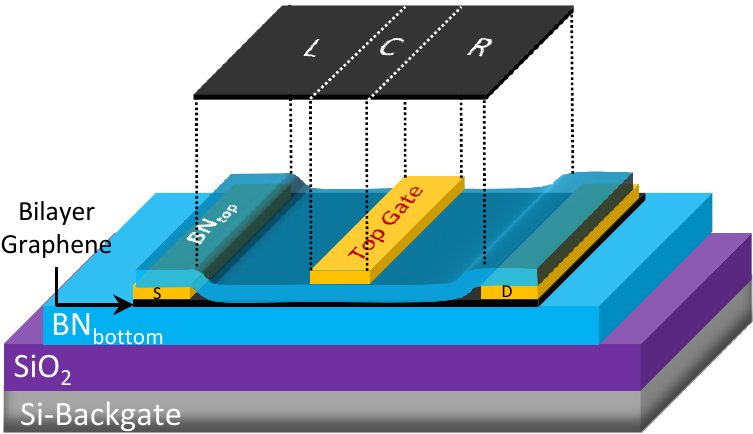}}

\subfloat[]{
\includegraphics[width=\columnwidth]{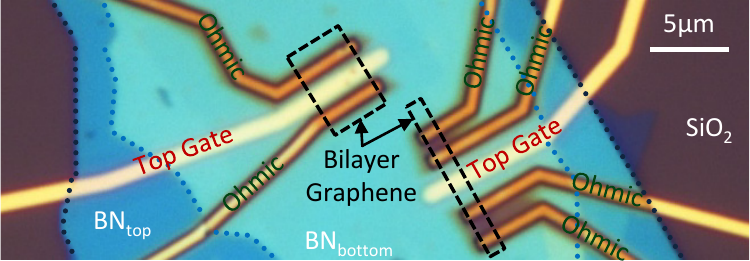}}
\end{center}
\caption{(a) Schematic of the device: a BLG flake is sandwiched between two h-BN layers. The whole flake can be tuned by the silicon backgate and the central area can additionally be independently tuned using the central top gate. Source (S) and drain (D) are Ohmic contacts enabling the measurement of the conductance $G$. The black plane at the top is a schematic of the three regions defined by the gates: L and R are the two leads and C is the dual-gated region. (b) Optical microscope image of two devices fabricated out of the same BLG flake. The dark yellow areas are Ohmic contacts (buried under $\mathrm{BN_{top}}$) and the bright yellow ones are the top gates. The measurements were carried out on the right device, using the inner Ohmic contacts. Adapted from \cite{varlet_fabry_2014}.}
\label{Exp_fig1}
\end{figure}

\subsection{Gate-tunable BLG interferometer}

The device under investigation is sketched in Fig.~\ref{Exp_fig1}(a) and (b) (right device). It consists of a h-BN/BLG/h-BN stack which is deposited on a Si/SiO$_{2}$-substrate, prepared as described in Ref.~\cite{varlet_anomalous_2014}. The stack was realized using the dry transfer technique pioneered in Ref.~\cite{Dean2010}.

For such a geometry, the silicon backgate allows us to tune the whole BLG stripe, whereas the local top gate acts only on the central region underneath. The device therefore consists of three areas in series, as shown in Fig.~\ref{Exp_fig1}(a): the two outer regions (labeled L and R) are simultaneously tuned by the backgate voltage $V_\mathrm{BG}$, and the central area (labeled C) is under the influence of both $V_\mathrm{BG}$ and  the topgate voltage $V_\mathrm{TG}$. The geometry is the following: the width of the whole flake is $W = 1.3\unit{\mu m}$, and the lengths of the three regions are $\ell_{\rm L}=\ell_{\rm R}=0.95\unit{\mu m}$ and $\ell_{\rm C}=1.1\unit{\mu m}$.

In the following, the sample is placed in a variable temperature inset at temperature $T = 1.6~\unit{K}$. We apply a constant symmetric bias voltage between the Ohmic contacts [labeled S and D in Fig.~\ref{Exp_fig1}(a)] and record the current, allowing to measure the conductance $G$. Additionally, the top gate voltage is modulated with a small AC voltage, which enables the measurement of the normalized transconductance $dG/dV_{\rm TG}$.

\subsection{Basic transport characterization}

\begin{figure}
\subfloat[]{
\includegraphics[scale=1.1]{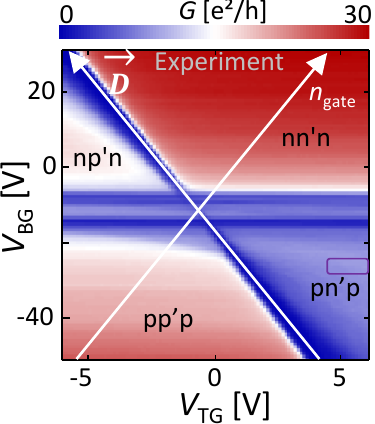}}
\subfloat[]{
\includegraphics[scale=1.1]{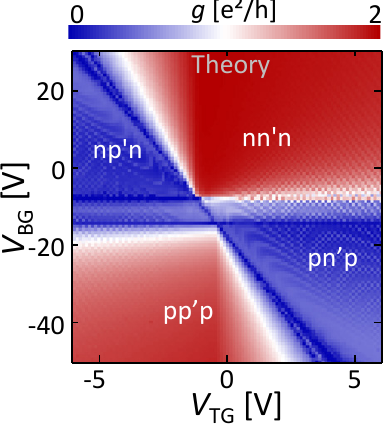}}
\caption{(a) Conductance versus top and backgate voltages \textit{measured} at $1.6~\unit{K}$. The purple box in the lower right corner indicates the range of the measurement shown in Fig.~\ref{Exp_fig2a}. (b) \textit{Calculated} normalized conductance versus top and backgate voltages. Adapted from \cite{varlet_fabry_2014}.}
\label{fig GVV and gVV}
\end{figure}

The measurement of such a dual-gated device is shown in Fig.~\ref{fig GVV and gVV}(a), where the conductance is displayed as a function of both $V_\mathrm{TG}$ and $V_\mathrm{BG}$. Depending on the applied voltages, the polarity of the outer regions and of the central one can be changed, from hole-like to electron-like transport. This gives rise to four different polarity combinations: two of them exhibit the same polarity in the three regions (\textit{pp'p} and \textit{nn'n} -- the prime referring to the central region C) and the other two exhibit opposite polarity (\textit{np'n} and \textit{pn'p}). The charge neutrality of the two outer regions is apparent from the two horizontal blue lines in the middle of Fig.~\ref{fig GVV and gVV}(a) and the charge neutrality in the dual-gated region occurs along the diagonal blue line. The latter spans along the so-called displacement field $\myvec{D}$ axis, as indicated in the figure. While increasing $|\myvec{D}|$, towards one corner of the map or the other, the conductance at the charge neutrality point decreases towards zero. This insulating state is due to the band gap being opened while the asymmetry between the top- and bottom layer of the BLG flake is made larger. Along the $\myvec{D}$-axis, the Fermi energy within the dual-gated region lies in the middle of the gap and the density is equal to zero. As reported in various experiments \cite{Oostinga2008,Russo2009,Weitz2010,Thiti_2010,Shimazaki2015,Sui2015}, hopping processes resulting from residual disorder were found to be the dominant transport mechanism in this regime and at such low temperatures.

In order to capture the electrostatic picture of the device and to reproduce our observations, we implemented the following model. First, the density $n$ in each of the three areas -- L, C and R -- was related to the applied gate voltages $V_\mathrm{TG}$ and $V_\mathrm{BG}$ via the use of a parallel plate capacitor model. Each area density is separated into three terms: the bottom graphene sheet density, the top graphene sheet density and the intrinsic doping of the area, which is directly estimated from the measured data. Following McCann \cite{mccann2013}, the asymmetry parameter $u$ is calculated in each region. With the knowledge of $u$ and $n$, the band offset is finally calculated and inserted in the nearest-neighbor tight-binding Hamiltonian of BLG~\cite{slonczewski1958,mcclure1957}. The details of the calculations are explained in Ref.~\cite{varlet_fabry_2014}. As a last step, the resulting energy profile is used to calculate the conductance through the device as a function of the applied gate voltages. This uses a Green's function formalism similar to Refs.~\cite{Liu2012a,rickhaus_ballistic_2013}. Here, only the C region is considered as the scattering region, L and R being treated as semi-infinite leads. As shown in Fig.~\ref{fig GVV and gVV}(b), which displays the normalized conductance $g$ (no mode counting implemented so that its maximum is $2e^2/h$ due to the valley degeneracy of the spin-independent calculation), the overall electrostatic picture is very well captured by the model.

\subsection{Interference pattern}
\begin{figure}
\subfloat[]{
\includegraphics[scale=1]{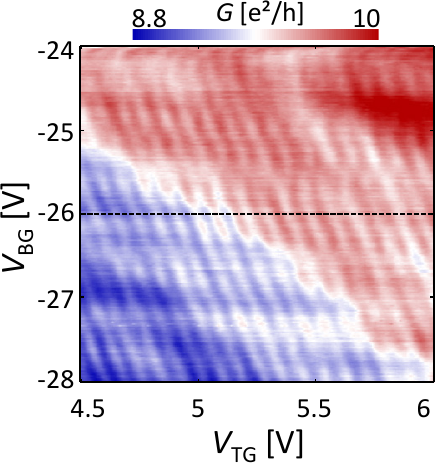}}
\subfloat[]{
\includegraphics[scale=1]{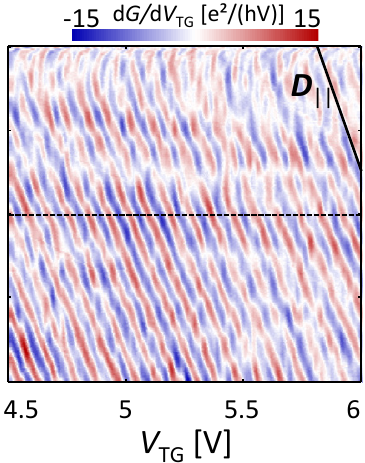}}

\subfloat[]{
\includegraphics[scale=1]{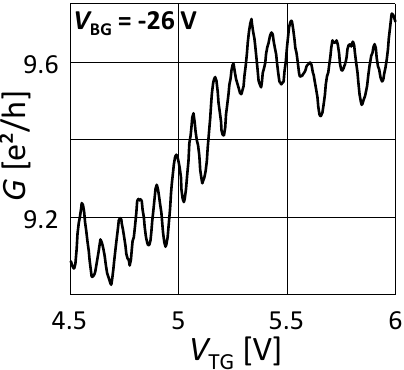}}
\subfloat[]{
\includegraphics[scale=1]{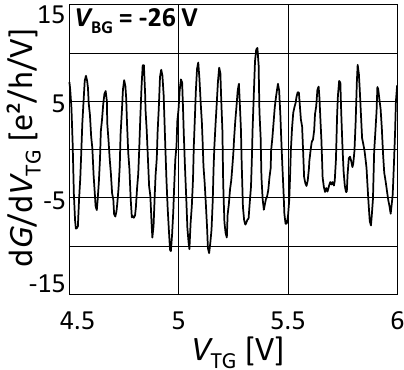}}
\caption{(a) Conductance measured as a function of top- and backgate voltages in the $pn'p$ regime [this region is highlighted with a purple rectangle in Fig.~\ref{fig GVV and gVV}(a)]. Conductance oscillations are clearly visible. (b) Normalized transconductance map, measured in the same range as in (a): the oscillations appear more clearly. (c)/(d) Cuts along the top gate voltage axis taken from (a)/(b) (black lines) at $V_{\mathrm{BG}}=-26~\mathrm{{V}}$. Adapted from \cite{varlet_fabry_2014}.}
\label{Exp_fig2a}
\end{figure}

As shown in Fig.~\ref{Exp_fig2a}(a), an oscillating conductance is observed in the \textit{pn'p} bipolar regime. To increase the visibility of the oscillations, one looks at the corresponding transconductance map, recorded simultaneously and shown in Fig.~\ref{Exp_fig2a}(b). Cuts within these two maps are shown in Fig.\ \ref{Exp_fig2a}(c)--(d). They allow to better see the strength of the oscillating pattern. In the following, we focus on the transconductance signal to analyze the oscillatory pattern in more detail. However, the same study could be carried out utilizing the conductance.

As highlighted in Fig.~\ref{Exp_fig2a}(b), the oscillations evolve parallel to the $\myvec{D}$-axis (the slope of the $\myvec{D}$-axis is displayed as a black line labeled $\myvec{D}\|$). This indicates that they arise from a mechanism taking place in the dual-gated part of the device. To confirm the ballistic origin of the observed signal, the frequency of the oscillations was analyzed. To do so, each top gate voltage point was converted into a density and then into a wavevector. Performing a discrete Fourier transform, we found that the oscillation frequency was $\lambda = 2.2~\unit{\mu m}$. This confirmed that the main contribution to the phase arises from interference between a directly transmitted wave and a wave which is transmitted through the cavity after bouncing once back and forth in the cavity. This corresponds to a phase difference $\Delta\Phi = k\cdot 2\ell_{\rm C}$, with a frequency $\lambda = 2\ell_{\rm C} = 2.2~\unit{\mu m}$. With this frequency analysis, we convince ourselves that the observation is related to ballistic transport in the dual-gated region, leading to a clear FP interference pattern.

\begin{figure}
\subfloat[]{
\includegraphics[width=\columnwidth]{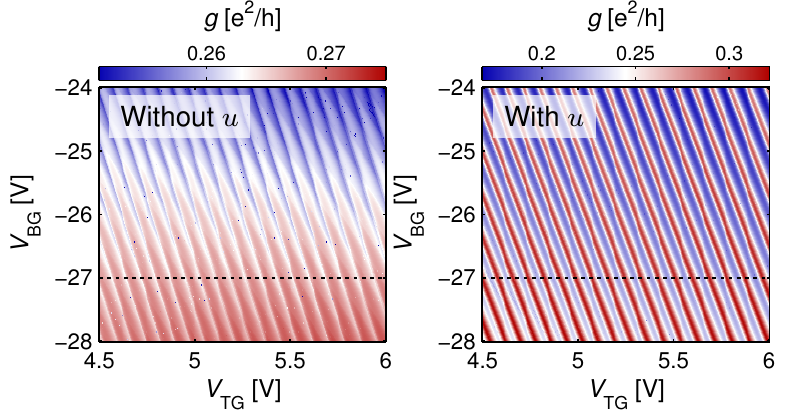}}

\subfloat[]{
\includegraphics[width=\columnwidth]{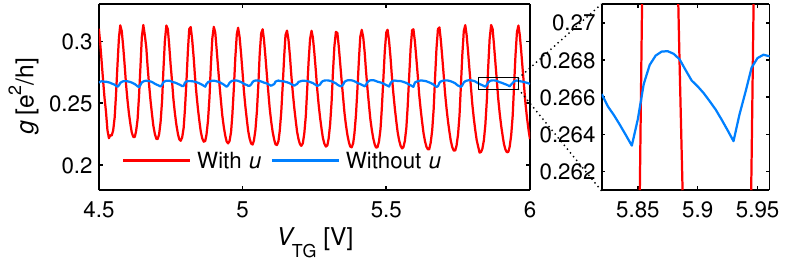}}
\caption{(a) Calculated normalized conductance as a function of top- and backgate voltages in the case for gapless BLG (left) and gapped BLG (right) (partly adapted from \cite{varlet_fabry_2014}). The latter considers the gate-tunable asymmetry $u$ following Ref.~\cite{mccann2013}. (b) A line trace cut from the maps of (a) indicated by the dashed lines therein.}
\label{BrokenOrNot}
\end{figure}

\subsection{Broken or not broken?}

From the previous analysis, a question remains to be explored: does the band gap play a role in our experimental observation? To elucidate this question, we first compare the measured oscillations shown in Fig.~\ref{Exp_fig2a}(b) with the theoretical predictions for two different systems: on one hand, a BLG system where the asymmetry between the layer is ignored is considered (i.e. the process of band gap opening is ignored and the gates are only responsible for tuning the overall sheet density) and, on the other hand, the same calculation method as already implemented in Fig.~\ref{fig GVV and gVV}(b) is used, which follows McCann's model \cite{mccann2013} and takes into account the asymmetry. The results of both calculations are shown in Fig.~\ref{BrokenOrNot}(a)--(b). Qualitatively, we observe that the shape of the oscillations is very different: in Fig.~\ref{BrokenOrNot}(a) (left panel), the oscillations appear very sharp and asymmetric, which is not the case in right panel, where the oscillations appear more symmetric, almost like a sine function [also visible in Fig.~\ref{BrokenOrNot}(b)]. Similar observations were made in Ref.~\cite{Rudner_2011}. Comparing these maps with the measurement data shown in Fig.~\ref{Exp_fig2a}(b), we conclude that our observation is \textit{qualitatively} closer to the case where the band gap is implemented.

Another confirmation lies in the visibility of the interference pattern. The visibility of the measured conductance oscillations is found to be $v = \Delta G/G_\mathrm{mean} = 1.5\%$. In a fully ballistic situation as those shown in Fig.~\ref{BrokenOrNot}(a)--(b), the visibility of the calculated conductance yields $v = 38\%$ for the case where $u$ is implemented and $v = 1.5\%$ when $u$ is set to zero. However, the assumption of a fully ballistic device has to be weakened. Indeed, by comparing the maps shown in Fig.~\ref{Exp_fig2a} and Fig.~\ref{BrokenOrNot}, one can see that the measurement signal exhibits imperfections. We would therefore expect the theoretical visibility values to be upper bounds compared to the experimental signal. We therefore conclude that the value $v = 1.5\%$ provided by the non-gapped case is inappropriate to describe our observation, pointing towards a role played by the band gap on our ballistic interference signatures.

\subsection{Gate-tunable Berry phase}
As explained in Section~\ref{section_3}, signatures of the broken chirality due to band gap opening should appear in the Berry phase, which varies as a function of the induced asymmetry and might be accessible in experiments. To find this out, the oscillations have to be investigated at varying magnetic field.

Based on the phase difference between a transmitted and a twice reflected electron wave, the FP resonance condition is
\begin{align}
\Phi_{\rm WKB}+\Phi_{\rm AB}+\Phi_{\rm Berry}=2{\pi}j,&& j\in\mathbb{Z}.
\label{res cond}
\end{align}
As explained earlier in Sec.~\ref{sec signatures of chirality}, $\Phi_\mathrm{WKB}$ and $\Phi_\mathrm{AB}$ are magnetic-field-dependent and are the origin of the parabolic trend of the oscillations evolving as a function of $B$. What strongly differs from the previously mentioned case of SLG is the Berry phase. In SLG, due to perfect transmission at normal incidence, the system requires finite magnetic field to build up trajectories which, in momentum space, enclose the origin and therefore pick up the Berry phase. In the case of gapped BLG, since even at $B=0$ the normal incident transmission is finite (in the studied energy range with finite $u$), there already exist trajectories which go through $k=0$, implying that the Berry phase is already involved.\footnote{Recall the trajectories sketched in Fig.~\ref{real and k trajectories without B}, which do not exist in the case of SLG due to the Klein tunneling. Here for gapped BLG, due to finite transmission and reflection at normal incidence, such trajectories do exist, so that the $k=0$ origin is always enclosed, independent of $B$.} The Berry phase therefore does not depend on the magnetic field \cite{varlet_fabry_2014}, but only on the asymmetry $u$.

\begin{figure}
\subfloat[]{
\includegraphics[height=4.7cm]{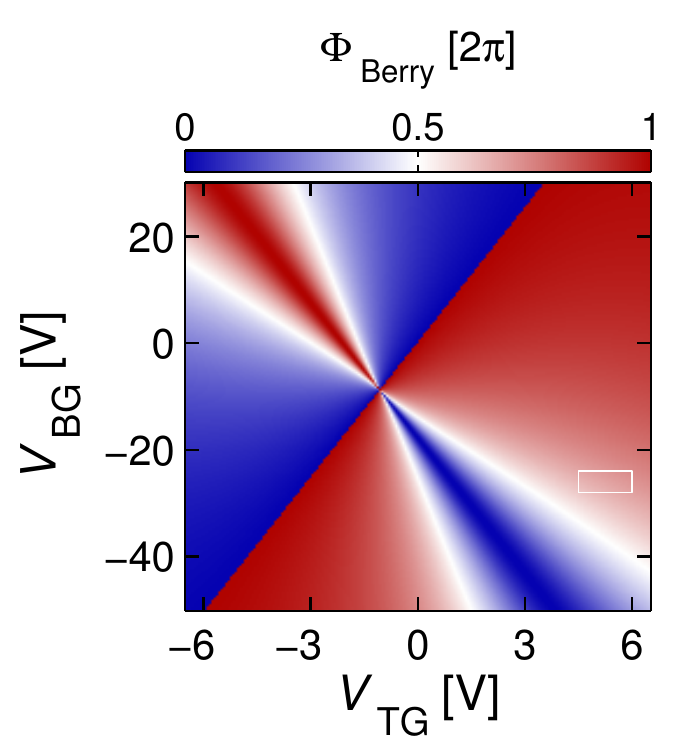}}\hfill
\subfloat[]{
\includegraphics[height=4.7cm]{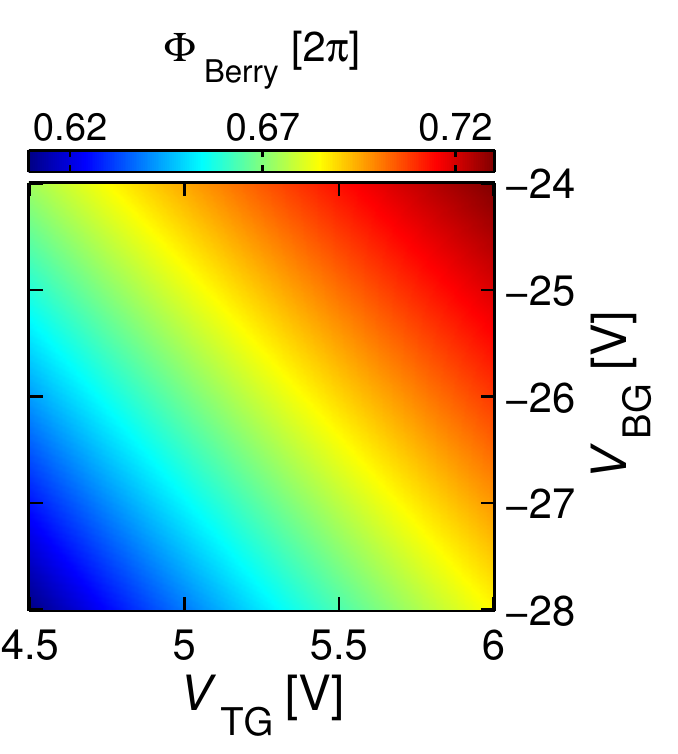}}
\caption{(a) Calculated Berry phase as a function of top- and back-gate voltages within the dual-gated area C of the device. (b) Zoom-in of the bipolar block indicated by the white box in (a). Adapted from \cite{varlet_fabry_2014}.}
\label{BerryExp}
\end{figure}

Figure \ref{BerryExp}(a) shows the predicted Berry phase of the dual-gated region of the device in the range of voltages available. We see that the Berry phase varies from $0$ to $2\pi$ in this map, and especially involves the value $\pi$ (white area), which is characteristic of SLG. However, since the interference is only visible in a limited range of gate voltage (white square marked in the map), we cannot probe the dramatic change of the Berry phase value. The values taken by the Berry phase in the region of interest are shown in Fig.~\ref{BerryExp}(b). We see that the Berry phase evolves, from $1.22\pi$ to $1.46\pi$ and is not constant. The limited evolution leads to only very small offsets of the oscillations as a function of top gate voltage, such that it is not visible in the experimental data. Further investigations would be required to probe the effect of the Berry phase, ideally with oscillations visible in a broader range of gate voltages.

Moreover, turning a BLG $pn'p$ into a SLG-like $pn'p$ by manipulating the gap and hence the Berry phase is theoretically possible, as shown in Section~\ref{section_3}. This would however require a system with three cavities, each being dual-gated. Additionally, the probing regime would have to cover the Berry phase of $\pi$ (which is supposed to be close to the gap). This would constitute one big step beyond the above-presented experiment where only one dual-gated region was designed.

\section{Conclusion and outlook: experiments based on the broken chirality}
We have explored theoretically the chirality of charge carriers both in SLG and BLG. We pointed out experimental realizations of FP interferometers which allow, in the SLG case, to directly probe the key parameter which is the Berry phase. We next focused on single- and bilayer graphene systems in which the inversion symmetry is lifted. Such systems exhibit band gaps in their band structure. Close to the edge of the gap, the pseudospin is found to be fully $z$-polarized, indicating a \textit{broken} chirality, which is then restored going away from the gap. This is associated with a Berry phase varying from zero to its pristine value ($\pi$ for SLG or $2\pi$ for BLG). Finally, we focused on the breaking of the chirality in BLG. Beyond the tunability of the anti-Klein tunneling, one of the key results is that the physics of BLG can mimic the behavior of SLG as the tunability of the transmission function allows to recover a perfect transmission at normal incidence for a certain energy range, together with a Berry phase of $\pi$, characteristic of the Klein tunneling in single-layer graphene. We presented an experiment on a dual-gated BLG graphene device where Fabry-P\'erot interference were probed, indicating ballistic transport. This allowed to illustrate the chirality breaking in more detail.

Very recently, experimentalists have been able to use the pseudospin degree of freedom both in SLG and in BLG. In 2014, Gorbachev et al.~\cite{Gorbachev2014} demonstrated the generation of topological currents in a SLG system with broken inversion symmetry. To do so, they aligned the graphene flake on a hexagonal boron nitride substrate, inducing an imbalance between the A and B atomic sites. Because graphene has two valleys, a broken inversion symmetry results in the creation of topological currents with different signs in each valley. This was confirmed by the observation of a non-local signal, as strong as the applied current and at micron distances from its path. The same was observed in a similar geometry in dual-gated BLG \cite{Shimazaki2015,Sui2015}. This time, dual-gating was responsible for breaking the inversion symmetry. A non-local resistance was measured, surviving up to high temperatures. Such advances in the control of the pseudospin degree of freedom are very appealing for possible applications in quantum computation, which require non-dissipative currents.

\section*{Aknowledgements}
We thank M.~Eich, H.~Overweg, and V.~Kr\"uckl for constructive comments and fruitful discussions. We also acknowledge financial support from the Marie Curie ITNs $S^3$NANO and QNET, together with the Swiss National Science Foundation via NCCR Quantum Science and Technology, the Graphene Flagship and the Deutsche Forschungsgemeinschaft within SFB 689.

\bibliographystyle{apsrev4-1}
\bibliography{pss_varlet_bib}

\end{document}